 \documentclass[useAMS,usegraphicx, usenatbib]{mn2e}
\usepackage{aas_macros}

\usepackage{amsmath}
\usepackage{amssymb}
\usepackage[utf8x]{inputenc}

\newcommand{\msun}{\,\text{M}_\odot}

\begin{document}
\title{The Effect of a Single Supernova Explosion on the Cuspy Density Profile of a Small-Mass Dark Matter Halo}

\author[R. S. de Souza,
	L. F. S. Rodrigues,
	E. E. O. Ishida and
	R. Opher]
{
R. S. de Souza$^{1,2}$\thanks{
Email: rafael@astro.iag.usp.br (RSdS);
felippe@astro.iag.usp.br (LFSR);
 emille@if.ufrj.br (EEOI);
 opher@astro.iag.usp.br (RO)
},
L. F. S. Rodrigues$^1$\footnotemark[1], 	
E. E. O. Ishida$^{1,2}$\footnotemark[1]
and
R. Opher$^1$\footnotemark[1]
\\
$^{1}$IAG, Universidade de S\~{a}o Paulo, Rua do Mat\~{a}o 1226, Cidade
Universit\'{a}ria,  05508-900, S\~{a}o Paulo, SP, Brazil\\ $^{2}$ IPMU,  
The University of Tokyo,
5-1-5 Kashiwanoha, Kashiwa, 277-8583, Japan}

 \date{Accepted -- Received  --}

\pagerange{\pageref{firstpage}--\pageref{lastpage}} \pubyear{2010}

\maketitle
\label{firstpage}

\begin{abstract}
Some observations of galaxies, and in particular dwarf galaxies, indicate a presence of cored density profiles in apparent contradiction with cusp profiles predicted by dark matter N-body simulations.
We constructed an analytical model,  using  particle distribution functions (DFs),  to show how a supernova (SN) explosion can transform a cusp density profile in a small-mass dark matter halo into a cored one.
Considering the fact that a SN efficiently removes matter from the centre of the first haloes, we  study the effect of mass removal through a SN perturbation in the DFs.
We found that the transformation from a cusp into a cored profile is present even for changes as small as 0.5\% of the  total energy of the halo, that can be produced by the expulsion of matter caused by a single SN explosion.
\end{abstract}
\begin{keywords}
cosmology: dark matter,  galaxies: haloes, supernovae: general
\end{keywords}

\section{Introduction}

 The presence of a cusp in the centre of  cold dark matter (CDM) haloes is one of the strongest results of N-body simulations \citep{mor99,NFW,nav04}.
However, the slope of this density profile is in apparent discrepancy with some observations of disk galaxies and galaxy clusters, which exhibit rather flat density cores \citep{bur95, kuzio2008,deb05,deb02, deblok2003,salucci2000}.  On the other hand,  e.g.,  \citet{coccato2008} studied the bulge and the disk kinematics of the giant low surface brightness galaxy ESO 323-G064 and showed that observations are not able to disentangle different density profiles, in that context. Parallel to the alternative models for dark matter already proposed, there are also attempts to solve this discrepancy within the CDM cosmology using baryonic physics (e.g., \citet{mas06,mas2008,governato2010}).

The so-called core/cusp problem has been in the spotlight of astrophysical research for quite a while (for a recent review, see \citet{deb10}). One natural attempt to solve the apparent contradiction,  is to study the gravitational effect of baryons in the dark matter density profile. This approach was first suggested by \citet{nav96}. The study of external impulsive mass loss events \citep{rea05} and steady winds \citep{gneezao} corroborated with the idea that baryons have a crucial role in the determination of the density profile \citep{ped09}. \citet{mas06} and \citet{pas10} discuss how star formation processes can interfere in the central profile shape.
\citet{pen10} argue that the existence of the central cusp is necessary in order to maintain the baryons trapped in the halo's gravitational potential. According to their results, the sum of SN explosions, tidal effects and star formation processes can completely destroy dwarf spheroidal haloes which initially present a core like central density.
Several authors suggested that the interstellar medium (ISM) of dwarf galaxy systems could be entirely removed by supernovae (SNe) explosions \citep{dek86,hen04, mor02,mor04,mor97, mur99, mac99}.
In this work, we study the effect of a single SN explosion on the removal of  baryon  gas from the first haloes and the effects of this removal on the shape of the density profile.

As stated in  \citet{mas06,mas2008} once removed, the cusp cannot be reintroduced during   subsequent mergers involved in  hierarchical evolution  of  galaxies. This statement is supported by numerical and analytical results. Outcomes from simulations made by  \citet{kazantzidis2006} imply that the  universal characteristic shape of dark matter density profiles may be set early in the evolution of halos and \citet{dehnen2005} shows theoretically that the remnant cusp cannot be steeper than any of the progenitor cusps.

In this work, we consider the possibility to solve the apparent core-cusp problem via baryonic physics \citep{mas06,mas2008}.  The sudden gas removal makes the dark matter distribution to expand. Such mechanism can operate efficiently in small high redshift haloes, creating dark matter cores \citep{governato2010}. Using arguments similar to those presented by
 \citet{tonini2006,desouza2010},  we analyze this process in the context of distribution functions.  
 
 It is initially assumed that baryons and dark matter are in an  equilibrium configuration.  Considering that energy perturbations  track the initial energy distribution (dominated mainly by dark matter), after being perturbed,  the halo will redistribute  dark matter particles in a new equilibrium configuration. As a consequence, variations in the initial  halo distribution function,  caused by energy or angular momentum perturbations,  are followed by macroscopic quantities, such  as the energy density profile.

In this paper,  we considered  the $\Lambda$CDM model with  best fit parameters from   \citet{jarosik2010} (WMAP-Yr7\footnote{http://lambda.gsfc.nasa.gov/product/map/current/}),  $\Omega_m = 0.267, \Omega_{\Lambda} = 0.734, \Omega_b = 0.0449$,   and $H_0 = 0.71$.
The paper is organized as follows.  In section \ref{sec:SNe},  we briefly review our knowledge about the propagation of blastwaves in the interstellar medium (ISM) and show the fraction of gas that can be removed from the haloes; in section \ref{sec:fd},   discuss the changes in density profile due to gas removal; and in section \ref{sec:results},  present and discuss our results. Finally, in section \ref{sec:conclusions},  we discuss our conclusions.

\section{Expulsion of baryonic  gas by a supernova explosion}	
\label{sec:SNe}

Since the  gravitational potential of the first collapsed haloes is shallow,  the ionizing radiation from the  first stars can expel the gas out of them \citep{abe07,alv06,kit04,wha04,wis08}. As a result, a subsequent SN can break away from the halo due to the decreased gas density by photoionization prior to the explosion \citep{kit05, wha08b}. In order to understand the nature of the SN shock expanding into an essentially uniform and ionized intergalactic space, we assume spherical symmetry.

\subsection{Evolution of a supernova blastwave}

As described in the review on astrophysical blastwaves by \citet{ost88}, and in the recent paper of \citet{sak09}, the evolution of a supernova remnant (SNR) in the intergalactic medium (IGM)  can be described as follows. Initially, the energy is largely thermal, but as the supernova expands, the adiabatic expansion converts  thermal  into kinetic energy.

During the  first stage, the SN ejecta sweeps out roughly the same amount of mass as its own in the surrounding medium.
 In the second stage, the expansion of the shock front is well approximated  by the Sedov-Taylor solution.  Eventually, radiative losses from the SNR become significant, and the remnant enters in the third, radiative, stage of its evolution. A thin shell is formed just behind the shock front. Finally the SNR expands conserving momentum.
 
\citet{vas08} found that supernova explosions with an energy $10^{53}$ ergs expel a significant portion of the initial baryonic mass from protogalaxies with total mass $\sim 10^{7} \msun$.  \citet{wha08} performed numerical simulations of primordial supernovae in cosmological haloes from $6.9\times 10^5-1.2\times 10^7 \msun$ and showed that even less energetic explosions are capable of ejection of more than $90\%$ of the baryons from haloes containing $\lesssim 10^7 \msun$. Based on such studies, we assume that feedback from SN explosions are able to expel up to all the baryonic gas in primordial haloes. Our next step then, is to evaluate what fraction of the total halo mass is in form of gas.

\subsection{Gas mass fraction}

In order to evaluate the ratio of ejected gas mass to total (virial) halo mass, we used the expression obtained by
\citet{Gnedin2000a}
\begin{equation}
f_{gas}(M,z)=\frac{\Omega_b/\Omega_m}{\left(1+0.26\frac{M_F(z)}{M}\right)},
\end{equation}
where $f_{gas}$ is the mass fraction of the halo, $M_F$ is the filter mass (e.g., \citet{gnedin2000,rodrigues2010,desouza2011}), $\Omega_b$ and $\Omega_m$ are the baryonic and dark matter energy density parameter, respectively and $M$ corresponds to the total (virial) mass of the halo. According to  \citet{Kravtsov2004}, $M_F\approx 5\times 10^6\msun$ at redshift $z=10$,  and the  reionization epoch is assumed to occur between $z \sim 8-11$ (e.g., the gas fraction of haloes with $M\sim 10^7 \msun$ was $f_{gas}=0.148$ at $z\sim 10$).

Since it is unlikely that the ejection of baryons was completely efficient, we consider cases where 5\%, 10\% and 15\% of the total mass was expelled, which correspond to 31\%, 61\% and  92\% of  expulsion of the halo's baryonic mass,  respectively.

\section{Evolution of the density profile of dark haloes}
\label{sec:fd}

In order to understand the effect on the density profile, caused by the removal of baryonic matter from the halo, we study the evolution of its distribution function (hereafter DF).

\subsection{Distribution functions and density profiles}

The DF fully describes the state of any collisionless system at any time, specifying the number of particles $f(x,v,t)\,d^{3}x\,d^{3}v$ having positions in the small volume $d^{3}x$ centered on $x$ and velocities in the small range $d^{3}v$ centered on $v$. The evolution of a collisionless system is governed by the Boltzmann equation, $\frac{d f}{d t}=0$ .

The DF of a  mass distribution in a steady state is related to its density profile through
\begin{equation}
\rho(r) = \int f(r,v)\,	d^{3}v\text{ .}
\end{equation}

For an isotropic galaxy, the density can be written in the simple form
\begin{equation}
 \rho(r) = 4\sqrt{2}\pi\int_{0}^{\Psi(r)}\left[\Psi(r)-\mathcal{E}\right]^{1/2}f(\mathcal{E})\,d\mathcal{E}\text{ ,}
 \label{eq:rho}
\end{equation}
where \( \Psi \)
is the relative potential and\[\mathcal{E} \equiv \Psi(r)-v^{2}/2 \] is the relative energy \citep{cud91,bin08}.

Equation (\ref{eq:rho}) is  an Abel integral whose solution is
\begin{equation}
 f(\mathcal{E}) = \frac{\sqrt{2}}{4\pi^2}\frac{d}{d\mathcal{E}}\int_{0}^{\mathcal{E}}\frac{d\rho}{d\Psi}\frac{d\Psi}{\sqrt{\mathcal{E}-\Psi}}\text{ .}
 \label{eq:fE}
\end{equation}

\subsection{Parametrizing the density profile}

The usual parametrizations of cuspy density profiles (eg. NFW) lead to potentials without a simple analytical solution for the inverse function, $r(\Psi)$. As a consequence, it is generally not possible to obtain an analytical form for the DF from equation (\ref{eq:fE}).

In order to avoid these complications,  we adopt the following family of spherical potentials \citep{rin09},
\begin{equation}
 \Psi = \frac{b^{\alpha\gamma}}{(b^{\alpha}+r^{\alpha})^{\gamma}} \text{ ,}
 \label{eq:Psi}
\end{equation}
which leads to density profiles of the form
\begin{equation}
\rho(r) = \rho_g  \alpha\gamma b^{\alpha\gamma} \frac{(1+\alpha)b^{\alpha}+(1-\alpha\gamma)r^{\alpha}}{r^{2-\alpha}(b^{\alpha}+r^{\alpha})^{\gamma+2}}\text{ ,}\label{eq:boaforma}
\end{equation}
where $\rho_g$ is a characteristic density to be adjusted for each profile,   
or, in terms of the relative potential
\begin{align}
\rho( \Psi)=&\frac{\rho_g }{1+\alpha} \Psi^{1+\frac{1}{\gamma}}
\left( \Psi^{-\frac{1}{\gamma}}-1\right)^{1-\frac{2}{\alpha}}\\
& \times\left[ 1-\alpha\gamma +\alpha(1+\gamma)\Psi^\frac{1}{\gamma}\right].\nonumber
\end{align}

This form of the density profile can conveniently capture the behavior of cusp and cored density profiles, for specific choices of the parameters $(\alpha, \gamma,b)$.

\subsubsection{NFW profile}
The most common choice for the form of the density profile of haloes, which is also in best agreement  with DM N-body simulations, is the NFW density profile \citep{NFW},
\begin{equation}
 \rho(r)=\frac{\rho_s}{\frac{r}{r_s}\left(1+\frac{r}{r_s}\right)^2}.
\end{equation}
Where $r_s$ and $\rho_s$ are determined by the concentration parameter $c$ through
\begin{equation}
 r_s=\frac{r_\text{vir}}{c}
\text{ and }
 \rho_s=\frac{\rho_\text{vir}}{3} \frac{c^3}{\log(1 + c) - \frac{c}{1 + c}}\text{,}
\end{equation}
where $\rho_\text{vir}= 178 \, \bar \rho$, is the average density of a virialized halo, and $r_\text{vir}$ corresponds to its radius.

Since we are treating very high redshift ($z > 10$) haloes (i.e. in the beginning of their mass accretion histories), we use $c=4$, following the prescription of \citet{Zhao2008}.

A different choice of $c$ should not, however, affect significantly the forthcoming results, since all of them can be expressed in function of the parameters $\rho_s$ and $r_s$ and will, thus, scale with a change in the concentration.

\begin{figure}\centering
  \includegraphics[width=0.9\columnwidth]{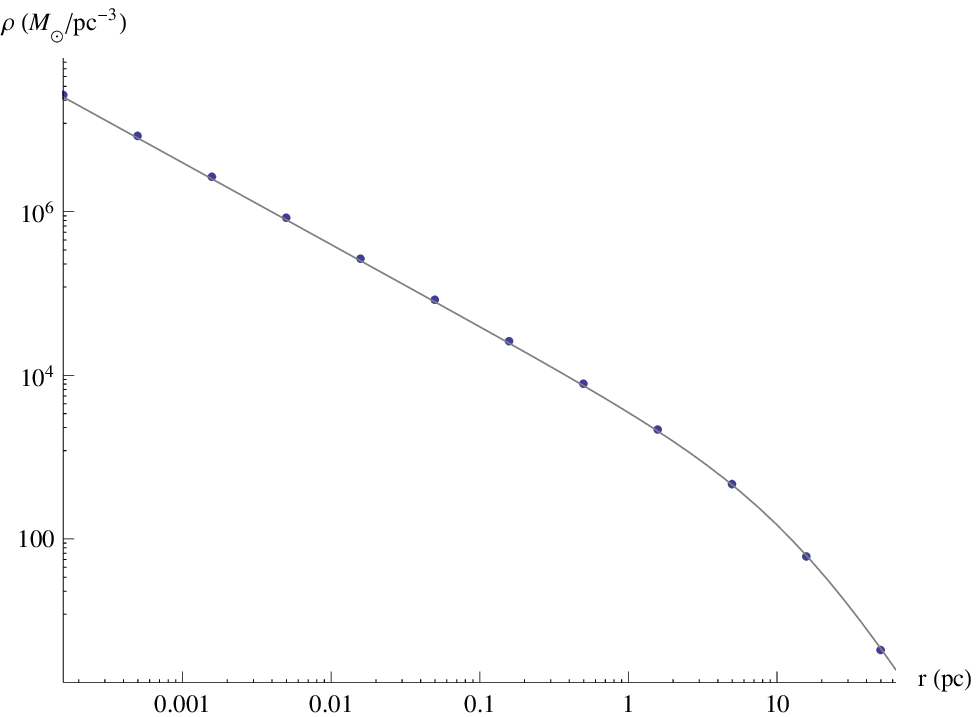}
  \includegraphics[width=0.9\columnwidth]{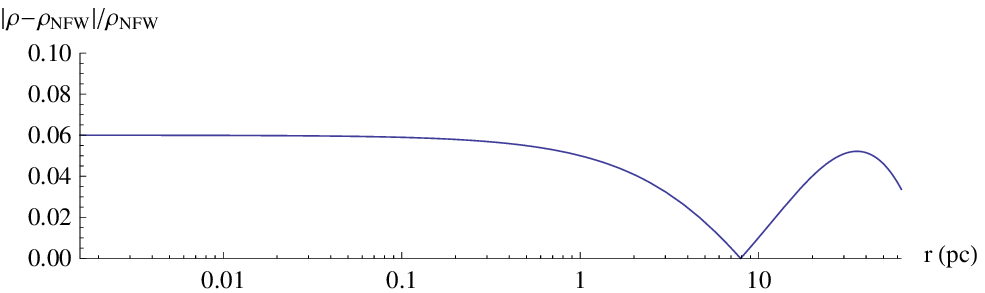}
  \caption{In the upper panel, the continuous (gray) curve shows the NFW density profile of a halo of mass $10^7\msun$ at $z=10$, and the (blue) points correspond to the general profile with $\alpha=1$, $\gamma=1/2$ and $b=1.085\,r_s=36.8\text{ pc}$. In the lower panel the fractional difference between the two profiles is shown.}
\label{fig:NFWadj}
\end{figure}

In figure \ref{fig:NFWadj} we show that the density profile of equation (\ref{eq:boaforma}) provides a very good approximation to the NFW profile for the choice $\alpha= 1$, $\gamma= \frac{1}{2}$, $b=1.085\, r_s$ and $\rho_g=\rho_s$, with the difference between the two profiles never exceeding 10\%. 

\subsubsection{Cored profiles}

Many recent observations \citep{bur95,Donato2009} favor  the Burkert density profile, which has the form
\begin{equation}
 \rho_\text{bur}(r)=\frac{\rho_0}{\left(1+\frac{r}{r_0}\right)\left(1+\frac{r^2}{r_0^2}\right)}\text{ .}
\end{equation}

\begin{figure}\centering
  \includegraphics[width=0.9\columnwidth]{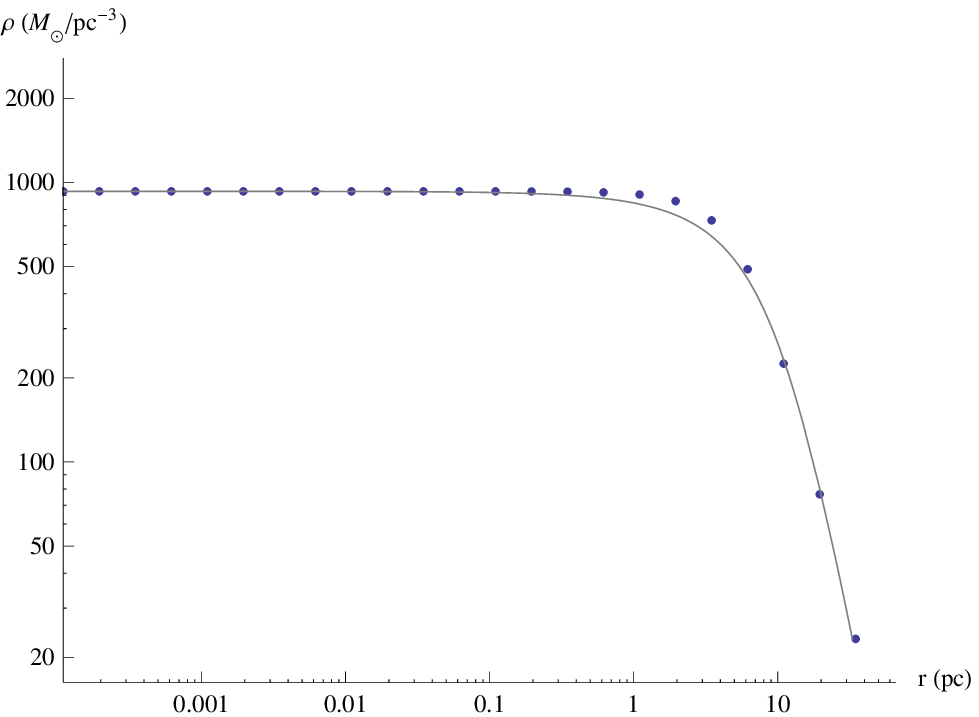}
  \includegraphics[width=0.9\columnwidth]{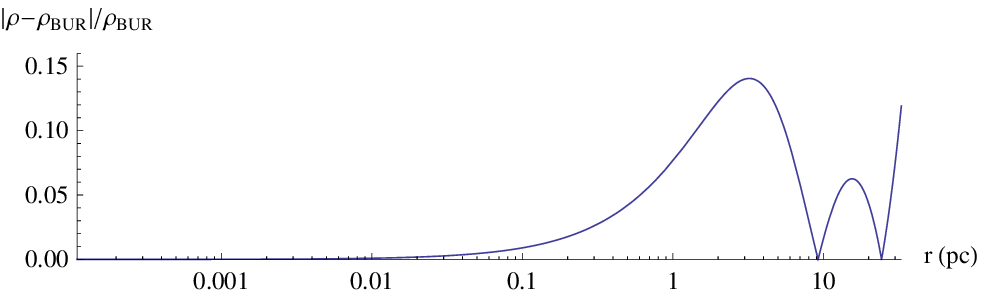}
  \caption{In the upper panel, the continuous (gray) curve shows the Burkert density profile of a halo of mass $10^7\msun$ at $z=10$, and the (blue) points correspond to the general profile with $\alpha=3/5$, $\gamma=6/5$ and $b=1.58\,r_s=53.7\text{ pc}$. In the lower panel the fractional difference between the two profiles is shown.}
\label{fig:BurAdj}
\end{figure}

The Burkert profile can be well approximated by equation (\ref{eq:boaforma}) choosing $\alpha= 2$, $\gamma= \frac{1}{16}$, $b=0.76\, r_0$ and $\rho_g=\rho_0$, as shown in figure \ref{fig:BurAdj}.  

Another common choice for a cored density profile, is the pseudo-isothermal (PI) profile
\begin{equation}\label{eq:PI}
 \rho_\text{iso}(r)=\frac{\rho_0}{1+\left(\frac{r}{r_0}\right)^2}\text{ .}
\end{equation}

\subsection{Perturbing the distribution function}
\label{sec:formalism}

The Boltzmann equation of a collisionless isotropic system is
\begin{equation}
 \frac{d\, f}{d\,t}=0 \quad\Rightarrow \quad\frac{\partial f}{\partial t}+\frac{\partial  \mathcal{E} }{\partial t}\frac{\partial f}{\partial  \mathcal{E} } =0\, \text{ .}
\end{equation}

The removal of the baryons from the halo will lead to a small change in its energy, $\delta  \mathcal{E} $. This change will take a small amount of time, $\delta t$, and will cause a change in the DF. We will  assume that the distribution function, after the perturbation, $f(t+\delta t, \mathcal{E})$, will have the form
\begin{equation}
 f(t+\delta t,  \mathcal{E} )\equiv  f(t, \mathcal{E} +\delta \mathcal{E})\,\, .\label{eq:removeE}
\end{equation}

To demonstrate the validity of the above assumption, we first expand the last  equation keeping only first order terms in $\delta E $,
\begin{equation}
  f(t+\delta t,\mathcal{E})\equiv  f(t,  \mathcal{E} +\delta \mathcal{E} )\approx f(t, \mathcal{E})+\delta \mathcal{E} \left.\frac{\partial f}{\partial{ \mathcal{E}}}\right|_{t,\mathcal{E} }\text{ .} \label{aeq:expandeDF}
\end{equation}

If the ejection took place on a small time interval, $\delta t$, the transformation in the distribution function can be written as
\begin{equation}
 \frac{\delta f}{\delta t}=\frac{f(t+\delta t,\mathcal{E})-f(t,\mathcal{E})}{\delta t}=-\left.\frac{\delta\mathcal{E}}{\delta t}\frac{\partial f}{\partial{\mathcal{E}}}\right|_\mathcal{E} ,
\end{equation}
which, in the limit of small time intervals, is precisely the Boltzmann equation.

\subsection{Relation between relative energy and perturbed halo mass}
The formalism discussed in section \ref{sec:formalism} allows us to understand how the removal from the halo, of a certain amount of relative energy, $\delta\mathcal{E}$, change the density profile.  It is, however, necessary to relate this change of  energy with the actual expulsion of baryons by  SNe.

From a fundamental point of view, this correspondence is far from trivial since the same amount of energy could be removed by either lower mass, faster particles or higher mass, slower particles. There is, also, a cosmological context, where taking into account the mass outside the virial radius is not a sensible choice.

We avoid those complexities taking the final density profile calculated  for a given $\delta\mathcal{E}$ and, then,  calculating the virial mass by a simple -- and fast convergent -- iterative procedure.

We initialize the virial radius variable, $r^{(0)}_{vir}$, with the virial radius of the unperturbed halo. We then calculate the virial mass using
\begin{equation}
 M^{(i)}_{vir}=\int^{r^{(i-1)}_{vir}}_0 4\pi r^2 \rho(r)\, dr
\end{equation}
where \(\rho(r)\) is the perturbed density profile. We reevaluate the virial radius
\begin{equation}
 r^{(i)}_{vir}= \left(\frac{M^{(i)}_{vir}}{\frac{4}{3} \pi \rho_{vir}} \right)^{1/3}
\end{equation}
and proceed to the next iteration.

This procedure converges rapidly to a consistent value for the virial mass of a perturbed halo, which can be compared with the known virial mass of the unperturbed halo (set to $10^7\msun$ in our calculations), therefore allowing us to relate variations in energy  with variations in mass.

\section{Results}
\label{sec:results}

Using the formalism developed in the end of section \ref{sec:fd}, we calculated the  DF associated with NFW density profile (the particular expression for the DF can be found in  appendix \ref{sec:calculatedDF}). The DF found were then perturbed through equation (\ref{eq:removeE}) and a transformed  density profile was generated from it using equation (\ref{eq:rho}).

\begin{figure}
\includegraphics[width=\columnwidth]{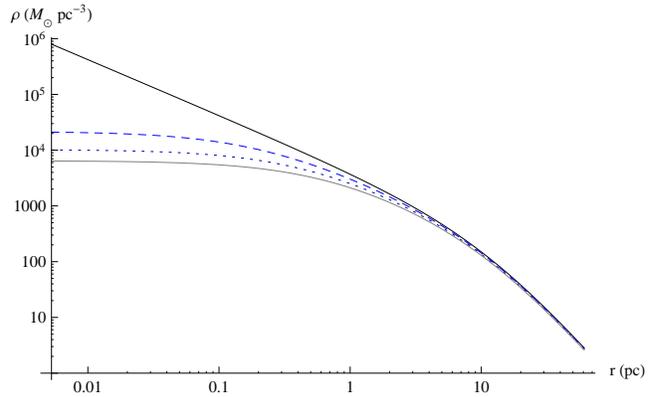}
  \caption{New equilibrium configuration of the perturbed halo removing 5\% (dashed curve), 10\% (dotted curve) or 15\% (thin curve) of the total mass of the halo,  compared to the NFW profile (top solid line).}
  \label{fig:densityNFW}
\end{figure}

In figure \ref{fig:densityNFW} we see how a NFW density profile varies due to the expulsion of 5\%, 10\% and 15\%  of mass of the halo (namely, removal of  
 0.57\%,  1.17\% and 1.81\%
 of the relative energy, respectively). The appearance of a core in the transformed density profile is noticeable.

In order to get better quantitative insight, we fit a pseudo-isothermal density profile to the resultant transformed profile. We found a core radii of  $r_0=0.054\,r_s $, $r_0= 0.075\,r_s $ and $r_0= 0.092\,r_s$
 for the removal of 5\%, 10\% and 15\% of the the halo's mass, respectively.

We also fit a Burkert-like profile finding $r_0= 0.20\, r_s$, $r_0 = 0.24\, r_s $ and  $r_0 = 0.28\, r_s $  for 5\%, 10\% and 15\%   of removal of the  halo's mass.

As in the NFW case, we fit a pseudo-isothermal density profile to the resultant transformed profile. We found core radii of
$r_0= 0.011\,r_s $,
$r_0=  0.012\, r_s $ and
$r_0=  0.013\, r_s $
for the removal of 5\%, 10\% and 15\%  of mass of the halo (namely, removal of  0.70\%,  1.44\% and 2.22\% of the relative energy, respectively). The fit of a Burkert-like profiles leads to
$r_0=  0.074\, r_s$,
$r_0 =  0.069\, r_s $ and
$r_0 =  0.064\, r_s $
  for 5\%, 10\% and 15\%  of removal of the the halo's mass. Our core radius are approximately similar to the one found by \citet{gneezao}. In their analysis, they $r_0$ lies in the range $0 \lesssim r_0 \lesssim 0.2 r_s$, depending of the model for outflow.

\begin{figure}
\includegraphics[width=\columnwidth]{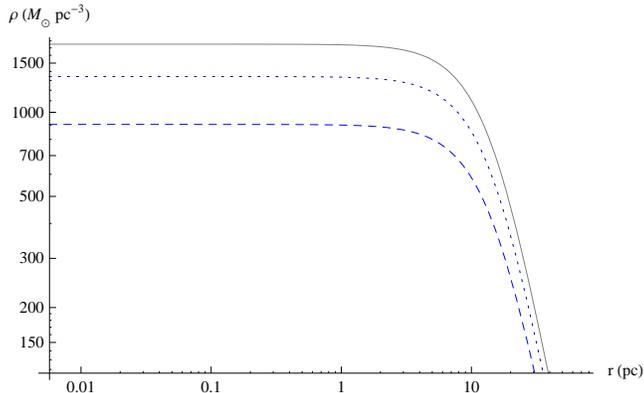}
  \caption{New equilibrium configuration of the perturbed halo removing 1\% (dotted line) and 2.5\% (dashed line) of total energy E,  compared to the original Burkert profile (solid line).}
  \label{fig:densityBur}
\end{figure}

As the  SN could also occur in a halo which was originally cored (e.g.  haloes where the core was previously erased by the process exposed), we performed the same analysis for a halo with an initial core profile. In figure \ref{fig:densityBur} we show the evolution of a Burkert density profile after energy removal.  As we can see,  the core structure  is maintained after gas removal and its radius remains approximately unchanged.

\section{Conclusions and discussion}
\label{sec:conclusions}

We explored the effect of a supernova explosion on the central density of a small mass ($\sim 10^7\msun$) dark matter halo at high redshifts.

We first reviewed (section \ref{sec:SNe}) the evidence that supernovae can efficiently expel a large part of the baryonic mass from small haloes.

We then  built a simple analytical model where we assume that
\begin{enumerate}
 \item the halo is approximately isotropic,
 \item the mass expelled by the SN  leads to a small loss of energy of the halo.
\end{enumerate}

Since typical parametrizations of the density profile do not lead to invertible expressions for the potentials, we used the profile of equation (\ref{eq:boaforma}), which is shown to be consistent with both cusp and cored density distributions.

From the density profile we calculated the distribution function associated with  NFW profile. We, then, evolved this distribution function, removing a small amount of energy  from it.

The final density profile found does not present a cusp, but, instead, a core.
We found that the transformation from a cusp into a cored profile is present even for changes as small as 0.5\% of the  total energy of the halo, that can be produced by the expulsion of matter caused by a single SN explosion.

\section*{Acknowledgements}

R.S.S. thanks the Brazilian agency FAPESP (2009/06770-2) and CNPq (200297/2010-4)  for financial support. 
L.F.S.R. thanks the Brazilian agency CNPq for financial support (142394/2006-8). E.E.O.I. thanks the Brazilian agency CAPES (1313-10-0) for financial support.
R.O. thanks the Brazilian agencies FAPESP (06/56213-9) and  CNPq (300414/82-0) for partial
support.  We also thank the anonymous referee for fruitful comments and suggestions.
\appendix

\section{ Distribution Function}
\label{sec:calculatedDF}
\subsection*{NFW distribution function}
The DF found for the choice of parameters $\alpha=1$,  $\gamma=\frac{1}{2}$ and $b=1.085 r_s$, that emulates a NFW density profile,  is
\begin{align}
 f(\mathcal{E})=& -\left.\frac{1}{56 \sqrt{2} \left(-1+\mathcal{E}^2\right)^3 \pi ^2}\right[ 4 (-1+\mathcal{E})  \mathcal{E}^{3/2} \nonumber \\
 & \times (1+\mathcal{E})\left(28-9 \,\mathcal{E}^2-136\,\mathcal{E}^4 96\, \mathcal{E}^6\right)\\\nonumber
 & +21 \sqrt{1-\mathcal{E}} \,(1+\mathcal{E})^3\,\text{arcsin}\left(\sqrt{\mathcal{E}}\right)\\\nonumber
& +21 (-1+\mathcal{E})^3 \sqrt{1+\mathcal{E}}\,\left. \text{arctanh}\left(\sqrt{\frac{\mathcal{E}}{1+\mathcal{E}}}\right)\right]\nonumber
\end{align}

\bibliographystyle{mn2e}

\label{lastpage}
\end{document}